\documentclass[a4paper,fleqn,usenatbib,useAMS]{mnras}
\usepackage{graphicx}	
\usepackage{amsmath}	
\usepackage{amssymb}	
\usepackage{multicol}        
\usepackage{bm}		
\usepackage{pdflscape}	

\usepackage[T1]{fontenc}
\usepackage{ae,aecompl}

\usepackage{mathptmx}
\usepackage{txfonts}

\title[Dust covering factors in RL \& RQ quasars]{Covering factors of the dusty obscurers in radio-loud and radio-quiet quasars}

\author[Gupta et al.]{
Maitrayee Gupta\thanks{E-mail: mgupta@camk.edu.pl},
Marek Sikora\thanks{E-mail: sikora@camk.edu.pl},
Krzysztof Nalewajko\thanks{E-mail: knalew@camk.edu.pl}
\\
Nicolaus Copernicus Astronomical Center, Bartycka 18,   00-716 Warsaw, Poland
}

\date{Last updated 2016 March 11; in original form 2016 March 11}

\pubyear{2016}

\begin{document}
\label{firstpage}
\pagerange{\pageref{firstpage}--\pageref{lastpage}}
\maketitle

\begin{abstract}
We compare covering factors of circumnuclear dusty obscurers in radio-loud and
radio-quiet quasars. The radio-loud quasars are represented by  
a sample of FR~II quasars obtained  by cross-matching a catalog of 
the FR~II radio sources selected by van Velzen et al. with the SDSS DR7 catalog 
of quasars.
Covering factors of FR~II quasars are compared with covering factors of 
the radio-quiet quasars matched with them in redshift, black hole mass, and Eddington-ratio.
We found that covering factors, proxied by  
the infrared-to-bolometric luminosity ratio, are on average slightly smaller
in FR~II quasars than in  radio-quiet quasars, however, this difference is statistically significant only for the highest Eddington ratios.
For both samples, no statistically significant dependence of 
a median covering factor on Eddington ratio, black hole mass, nor redshift  can be 
claimed.
\end{abstract}


\begingroup
\let\clearpage\relax
\endgroup
\newpage

\section{Introduction}

Discovery of broad lines and strong non-stellar continuum in polarized light of Seyfert 2 galaxies \citep{1985ApJ...297..621A,1987BAAS...19..695M}
and thermal infrared (IR) excesses \citep{1985ApJ...293L..11M,1987ApJ...321..233E}
led to idea about existence in active galactic nuclei (AGN) of dusty, molecular tori \citep{1988ApJ...329..702K}.
This allowed us to unify
Seyfert~2 galaxies (only narrow lines and weak or absent non-stellar continuum)
with Seyfert~1 galaxies (broad emission lines and strong non-stellar continuum).
The dusty tori were proposed also to unify narrow-line radio galaxies with radio-loud (RL) quasars and broad-line radio galaxies \citep{1989ApJ...336..606B},
and discovery of radio-quiet (RQ) quasars of type~2 (\citealt{2008AJ....136.2373R} and refs. therein) confirmed the presence of dusty obscurers in RQ quasars as well.

As indicated by the number ratio of the type~2 AGN to the type~1 AGN, and, independently, by the infrared-to-UV luminosity ratio, these tori are geometrically thick \citep{2015ARA&A..53..365N}.
They are predicted
to have inner edge at the sublimation radius (\citealt{1969Natur.223..788R}; \citealt{1987ApJ...320..537B}) and are
likely enclosed within the black hole (BH) influence sphere (\citealt{2012MNRAS.420..320H} and refs. therein).
Location of the inner edge at the sublimation radius is confirmed 
by reverberation mappings (e.g. \citealt{2014ApJ...788..159K}) and by  near-IR interferometry  \citep{2012JPhCS.372a2033K}, while constraints on their spatial extensions are suggested  
by the mid-IR interferometry \citep{2013A&A...558A.149B,2016A&A...591A..47L}.
Hence, studies of such tori provide excellent opportunity to probe the conditions of the outermost portions of the BH accretion flows.
Unfortunately, physical and dynamical structure of dusty tori is still not known (see review by \citealt{2015ARA&A..53..365N}).
It is even uncertain, whether they are really tori, or maybe the tori  are just mimicked by  optically thick portions of disc winds (e.g. \citealt{1992ApJ...385..460E,1994ApJ...434..446K,2006ApJ...648L.101E}) or by warped, tilted accretion discs (\citealt{2010ApJ...714..561L} and refs. therein).
One might try to resolve the real structure of the dusty region by modelling spectra of IR radiation resulting from reprocessing of optical/UV central radiation (see e.g. \citealt{2016MNRAS.458.2288S}).
However, due to the complexity of these models, the observed spectra can be  reproduced by many of them 
with a broad choice of the adopted model parameters \citep{2016ApJ...819..123N}.

Regardless of the uncertainties in the `torus' structure, using the mid-IR spectra one
can at least study the fraction of the central radiation that is reprocessed by circumnuclear dust ($R \equiv L_{\rm MIR}/L_{\rm bol}$), in particular its distribution and dependence on the Eddington ratio ($\lambda_{\rm E} \equiv L_{\rm bol}/L_{\rm Edd}$) and black hole mass ($M_{\rm BH}$).
Noting that luminous RQ and RL quasars cover similar ranges of these parameters, by
comparing above properties in these two quasar populations  one may try to verify whether 
the conditions in their outer accretion discs are different, as one might expect in case of
having accretion triggered by mergers in RL quasars and by secular processes in RQ quasars \citep{2015ApJ...806..147C}, and/or having accretion flows characterized by higher magnetization in the RL quasars as compared with the RQ ones.
Obviously, if existing, the largest difference in
the $R$ should be noticed between the RQ quasars and the most RL ones.
Hence, unlike in similar studies
performed by \cite{2013MNRAS.430.3445M}, our RL quasar sample was chosen to be represented only by quasars with Fanaroff-Riley Class II (FR II) radio-morphology.
Furthermore, in order to exclude
possible differentiating effects coming from different BH masses, Eddington-ratios, and redshifts,
we perform such comparison by respective pairings of RL and RQ quasars.
First, when deriving  dependence of $R$ on $\lambda_{\rm E}$ , we pair each RL quasar with a RQ one with the same redshift and black hole masses.
Secondly, when studying dependence of $R$ on the black hole mass, we pair quasars with the same redshift and $\lambda_{\rm E}$.
Thirdly, when studying dependence of $R$ on the redshift, we pair quasars with the same $\lambda_{\rm E}$ and black hole mass.

Our work is organized as follows: 
in \S \ref{sec_samples} we define our data samples, in \S \ref{sec_methods} the methods of calculation of $R$ and
pairing procedures are described, in \S \ref{sec_results} results of our studies of covering factors
of dusty obscurers in FR~II versus RQ quasars are presented, and in \S \ref{sec_discussion}  
they are discussed in the context of the radio diversity of quasars.
We adopted $\Lambda$ cold dark matter cosmology, 
with $H_0 = 70 \rm{km\,s}^{-1}$, $\Omega_m=0.3$ and $\Omega_\Lambda=0.7$.

\section{Data samples}
\label{sec_samples}

\subsection{FR II quasars}

A large catalogue of 59192 FR~II radio sources \citep{1974MNRAS.167P..31F} was recently compiled by \cite{vVel15}, using data from the Faint Images of the Radio Sky at Twenty-centimetres (FIRST) catalogue \citep{1995ApJ...450..559B}.
From this, \cite{vVel15} selected a sample of 1108 FR~II quasars by matching their FR~II radio sources with the combined DR7 and DR9 catalogues of SDSS quasars.
We are interested only in the DR7 quasars, for which black hole mass estimates are available \citep{She11}.
Since sources that are both in the DR7 and DR9 are denoted as `DR9', in order to select all DR7 quasars, we performed additional matching of the FR~II quasar sample of \cite{vVel15} (1108 sources) with the SDSS DR7 quasar catalogue \citep{2010AJ....139.2360S} (105783 sources).
We used a matching radius of 5 arcsec and obtained 899 objects.
This resulting sample of FR II quasars was then matched with the sample of SDSS DR7 quasars detected by the \textit{Wide-field Infrared Survey Explorer (WISE)} \citep{Wu12}.
This gave us 895 FR II quasars detected in the MIR band.

\subsection{Radio-quiet quasars}

The RQ sample with MIR data is constructed by matching the DR7 quasar catalogue \citep{2010AJ....139.2360S} and \textit{WISE} all-sky catalogue \citep{Wu12}, using a matching radius of 1 arcsec, resulting in 101853 objects.
From these we remove the 899 RL quasars matched with the catalog by \cite{vVel15}, this leaves us with 100958 quasars.
We then remove objects that were detected by the FIRST survey \citep{1995ApJ...450..559B}, this gives us 92648.
We repeat the same process with the NVSS \citep{1998AJ....115.1693C} and end up with 92445 objects.
We also removed those objects that were outside the FIRST observation region.

\subsection{Additional sample criteria}

As it is stated in \S3, we calculate the MID infrared luminosity,  using the flux measured in
the \textit{WISE}/W3 window ($\lambda \simeq 12\,{\rm{\mu}m}$). In order to avoid having this window  too close to the short-wavelength edge of the MIR band in the quasar rest frame,
we limit our samples to $z < 2$. 
We also introduce a limit on black hole masses, taking only quasars with $M_{\rm BH} > 10^8 M\odot$.
This is in order to avoid Seyfert galaxies,
in which the MIR flux can have a significant contribution from starbursts.
Furthermore, putting limits on redshift and black hole mass reduces
biases imposed on studied properties by selection effects related to the flux limits.
Finally, we constrain our samples to the objects with the Eddington-ratio $\lambda_E > 0.01$
to assure that accretion flows in the selected objects are radiatively efficient.
After applying all these limits we end-up with 797 FR~II quasars and 67480 RQ quasars.

\begin{figure}
\centering
\includegraphics[scale=0.5]{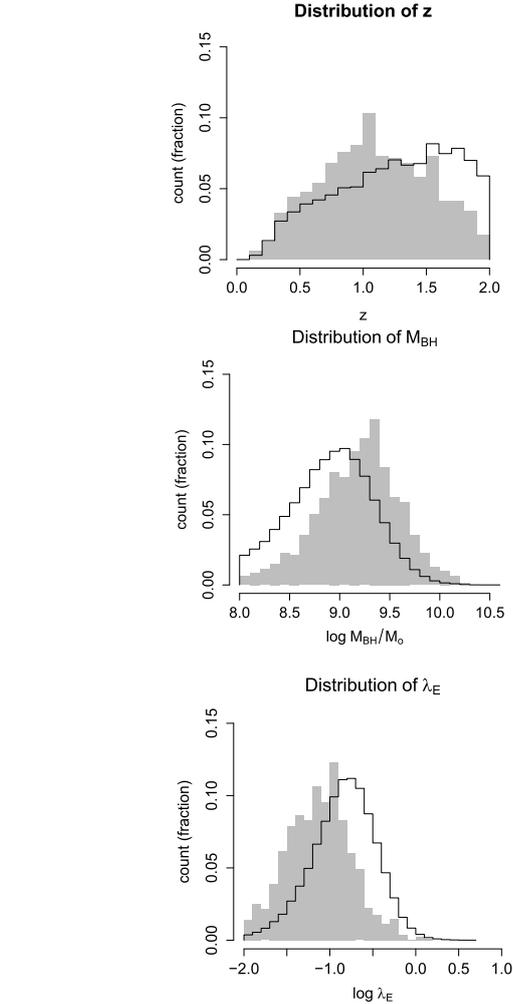}
\caption{Distributions of our RL (\emph{grey}) and RQ (\emph{unfilled}) samples of quasars in redshift $z$, black hole mass $M_{\rm BH}/M_\odot$, and Eddington radio $\lambda_E = L_{\rm bol}/L_{\rm Edd}$.}
\label{img_hist}
\end{figure}

In Figure \ref{img_hist} we show the histograms of our RL and RQ samples as a function of redshift $z$, black hole mass $M_{\rm BH}/M\odot$, and Eddington radio $\lambda_{\rm E}$.

\section{Methods}
\label{sec_methods}

\subsection{Covering factor}

There are two main methods to derive covering factors of dusty, circumnuclear obscurers, using number ratio of type 1 to type 2 AGNs or using ratio of the MIR luminosity to the bolometric luminosity (see review by \citealt{2015ARA&A..53..365N}). Both methods suffer from several drawbacks, the former because of difficulties to assure the common parent AGN population and because of having dependence of $N_2/N_1$ on the covering factor distributions \citep{2011ApJ...731...92R,2012ApJ...747L..33E},
the latter because derivation of the covering factor from $L_{\rm MIR}/L_{\rm bol}$ requires knowledge of anisotropy of accretion disc  and obscurer radiation, and those
are model dependent.
In this paper we will assume  that the  covering factor is simply equal to  $R \equiv L_{\rm MIR}/L_{\rm bol}$, which corresponds with the assumption of isotropic radiation of both components. While quantitatively this may 
lead to large systematic departures from real values of covering factors (see, e.g. \citealt{Cald12}), as long as our main interest is 
to verify whether dusty obscurers in RL and RQ quasars are different, using $R$ is not expected to affect trends studied  by us.  
We calculate $R$ taking $L_{\rm bol}$ from \cite{She11} and $L_{\rm MIR}$ using formula $L_{\rm MIR} = 12.5 \times (\nu_{W3} L_{\nu,W3})$,
where $\nu_{\rm W3} = 10^{13.4}$\,Hz, $L_{\rm\nu,W3} = 4 \pi d_L^2 F_{\rm\nu,W3}$,
$\log F_{\rm\nu,W3} = -(5.174 + m_{\rm W3})/2.5$,
and $m_{\rm W3}$ magnitudes are provided by \cite{Wu12},
denoted in Table~\ref{tbl_1} as `W3'.

\begin{table}
\caption{Center wavelengths and frequencies of the four WISE bands. The fourth column gives \( \Delta \)m which is used to transform the the given Vega magnitudes to the AB magnitude system.}
\label{tbl_1}
\begin{center}
 \begin{tabular}{c c c c} 
 \hline
 Band & $\lambda$($\mu$m) & log($\nu$) (Hz) & $\Delta$m \\ [0.5ex] 
 \hline\hline
 1 & 3.435 & 13.94 & 2.699 \\ 
 2 & 4.6 & 13.81 & 3.339 \\
 3 & 11.56 & 13.41 & 5.174 \\
 4 & 22.08 & 13.13 & 6.620 \\
 \hline
\end{tabular}
\end{center}
\end{table}

Our choice of using \textit{WISE}/W3 data to calculate $L_{\rm MIR}$ can be justified by noting that:

\begin{itemize}
\item
circumnuclear dust spectra in the AGN rest frames are approximately flat, i.e., with a spectral index $\alpha_{\rm MIR} \simeq 1$ ($\alpha: F_{\nu} \propto \nu^{-\alpha_{\rm MIR}}$) and enclosed between   
$2 {\mu}m$ and $25 {\mu}m$ \citep{2011ApJ...736...26H,2015ARA&A..53..365N};

\item
for redshifts $z > 1.3$ the W1 and  W2 bands move in  the quasar rest frame to $\lambda_{\rm E} < 2 {\mu}m$;

\item
W4 fluxes  have much larger S/N than in other channels, hence calculating $L_{\rm MIR}$ by using
spectral slopes between separated by less than factor of 2 $\nu_{\rm W2}$  and $\nu_{\rm W3}$ 
may create much larger errors than using only W3 flux and adopting $\alpha_{\rm MIR} = 1$.
\end{itemize}

\subsection{Pair Matching}
\label{sec:pair}

We use a matching technique used in \cite{Kauf08}, \cite{Mast10} and \cite{Wild10} instead of globally comparing the properties of the RL and RQ samples with different distributions of black hole mass or redshift.
In order to study the dependence of $R$ on any of the three parameters $\lambda_E$, $M_{\rm BH}$ or $z$, each RL quasar is associated with the same number $n$ of RQ quasars according to their matching distance calculated for the other two parameters.
First, candidate RQ counterparts are selected from the following range:  $|\Delta\log \lambda_E| < 0.09$, $|\Delta z| < 0.01$ and  $|\Delta \log M_{\rm BH}| < 0.12$.
Secondly, we calculate the matching distances between the RL quasar and the RQ candidates defined as $d_{\rm match}^2 = \sum_i \Delta_i^2$, where $\Delta_i$ stands for either $\Delta\log\lambda_{\rm E}$, $\Delta z$ or $\Delta\log(M_{\rm BH}/M\odot)$.
Then, we select only those RQ candidates which have the shortest matching distance $d_{\rm match}$.
We have considered different numbers of paired RQ quasars with $n = 1,3,10$.
This process of matching can result in the same RQ quasar paired with more than one RL ones.

Then, the sample of RL quasars, and the subsample of paired RQ quasars, are divided according to the studied parameter into several bins containing equal number of sources.
For example, in order to study the dependence of $R$ on $\lambda_{\rm E}$,  we divide the samples in the bins of $\lambda_{\rm E}$, and then we pair the quasars in $M_{\rm BH}$ and $z$ according to the matching distance given by $d_{\rm match}^2 = (\Delta z)^2 + [\Delta\log(M_{\rm BH}/M\odot)]^2$.

\begin{figure}
\centering
\includegraphics[scale=0.5]{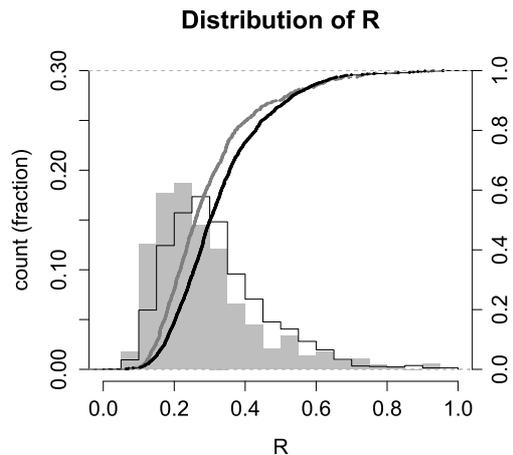}
\caption{Normalised and cumulative distributions of dust covering factor $R$ for our RL sample (\emph{grey}) and the RQ subsample paired with $n = 3$ (\emph{unfilled / black}).}
\label{img_hist_r}
\end{figure}

\section{Results}
\label{sec_results}

Figure \ref{img_hist_r} compares the overall normalized distributions of $R$ for RL and RQ quasars that are paired in $z$, $M_{\rm BH}$ and $\lambda_{\rm E}$ with $n = 3$.
We have overlaid the cumulative distributions of the two samples.
The typical values of $R$ for both RL and RQ quasars are in the range $0.1-0.6$. The histograms peak at $R_{\rm peak} \simeq 0.25$, with $R_{\rm peak}$ being slightly higher for the RQ quasars.
The mean values of $R$ are $0.28$ for RQ quasars and $0.32$ for RL quasars, and the median values are $0.25$ and $0.29$, respectively.

\begin{figure*}
\centering
\includegraphics[width=\textwidth]{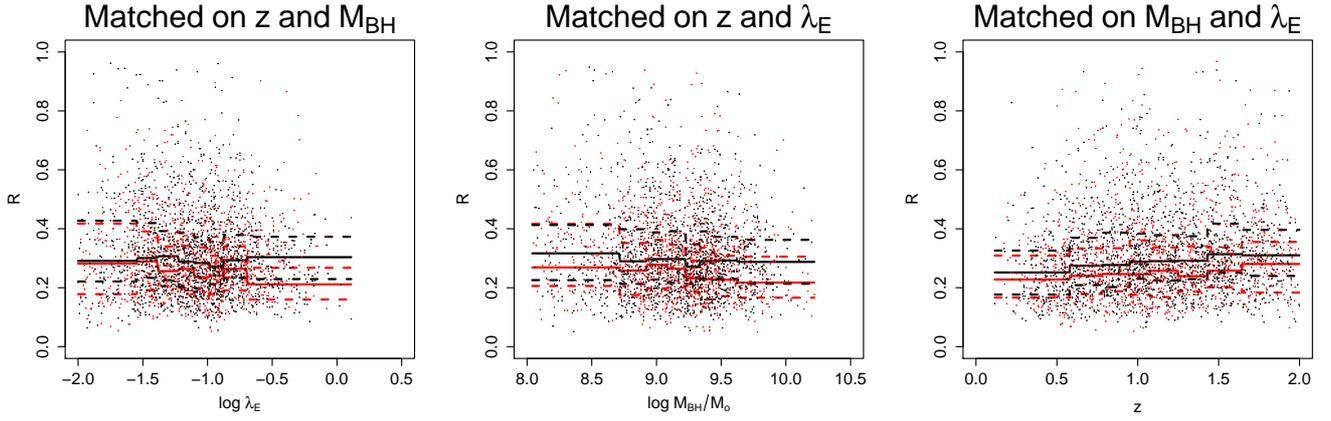}
\caption{Dependence of $R$ on Eddington ratio $\lambda_E$ shown in the \emph{left panel}, black hole mass $M_{\rm BH}$ shown in the \emph{middle panel} and redshift $z$ as shown in the panel to the \emph{right}.
The \emph{red points} represent the RL sample, while \emph{black points} represent the RQ subsample paired with $n = 3$.
The \emph{solid lines} show the median value for bins containing equal numbers of sources.
The \emph{dashed lines} represent the $25^{\rm th}$ and $75^{\rm th}$ percentiles of the respective data.}
\label{best_three}
\end{figure*}

\begin{figure*}
\centering
\includegraphics[width=\textwidth]{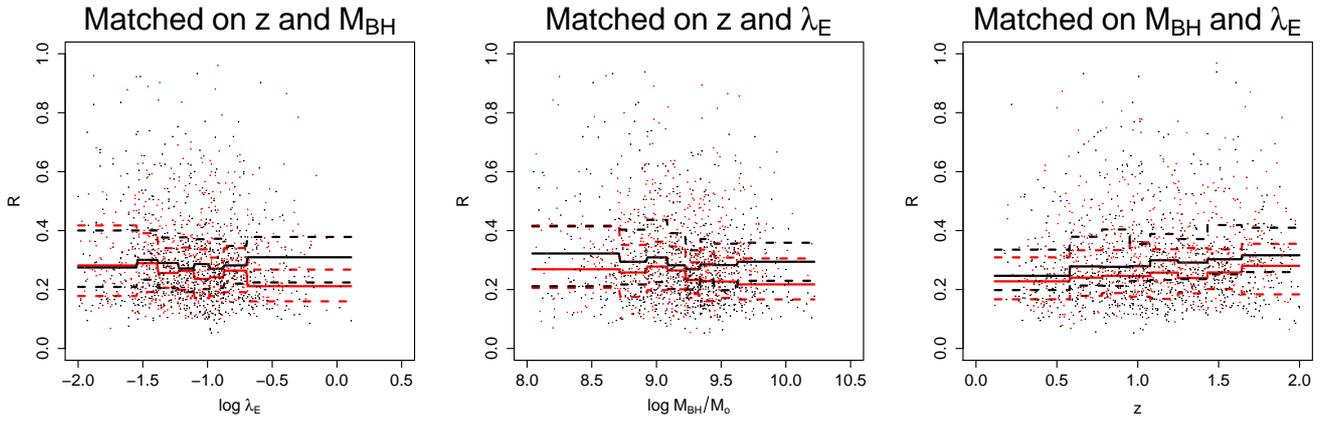}
\caption{Same as Figure \ref{best_three}, but for $n = 1$.}
\label{best_match}
\end{figure*}

In Figure \ref{best_three} and Figure \ref{best_match} we show the distributions of $R$ versus either $\lambda_{\rm E}$, $M_{\rm BH}$ or $z$ for the paired samples of RL and RQ quasars.

\begin{figure*}
\centering
\includegraphics[width=\textwidth]{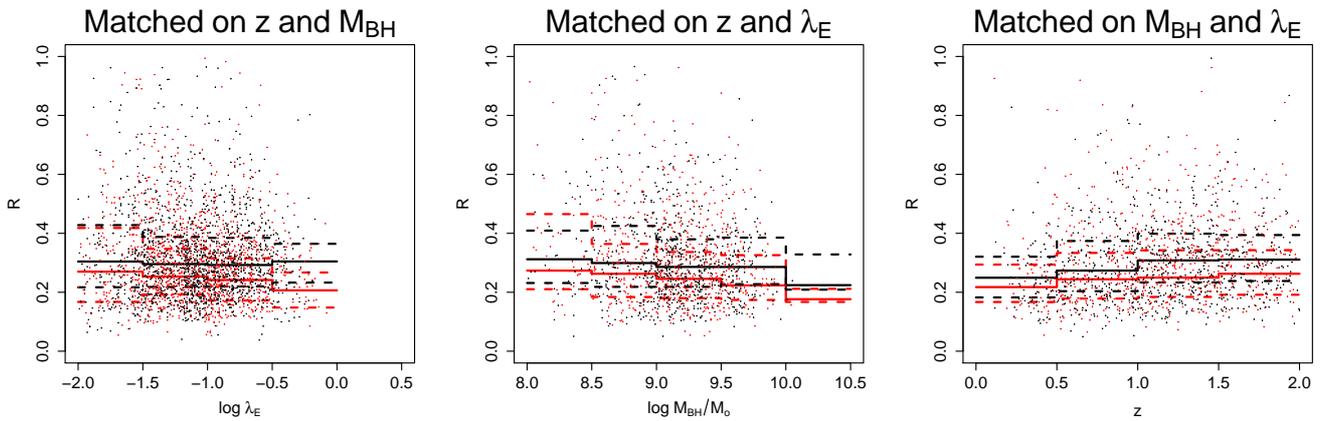}
\caption{Same as Figure \ref{best_three}, but with equal sized bins.}
\label{bin_match}
\end{figure*}

Table \ref{tbl_3} reports the median values $R_{50}$, as well as first and third quartiles $R_{25}$ and $R_{75}$ for different bins of $\lambda_{\rm E}$, corresponding to the left panel of Figure \ref{best_three}.
There is no overall trend that we see in the distribution of $R$.
The difference in $R$ between RL and RQ increases with the increase in $\lambda_E$, z, and $M_{\rm BH}$.
This trend remains the same, irrespective of the number of matches being $n = 1,3,10$, and whether the bins are equally spaced or contain equal number of objects. We added an additional plot Figure \ref{bin_match} with equal sized wide bins and we do not see any changes in the trend. 

We also observe that $R_{25}$ and $R_{75}$ indicate broad distributions of $R$ for every $\lambda_{\rm E}$, $M_{\rm BH}$ and $z$.
Therefore, variations in the median values do not appear to be significant.

\begin{figure*}
\centering
\includegraphics[width=\textwidth,height=\textheight,keepaspectratio]{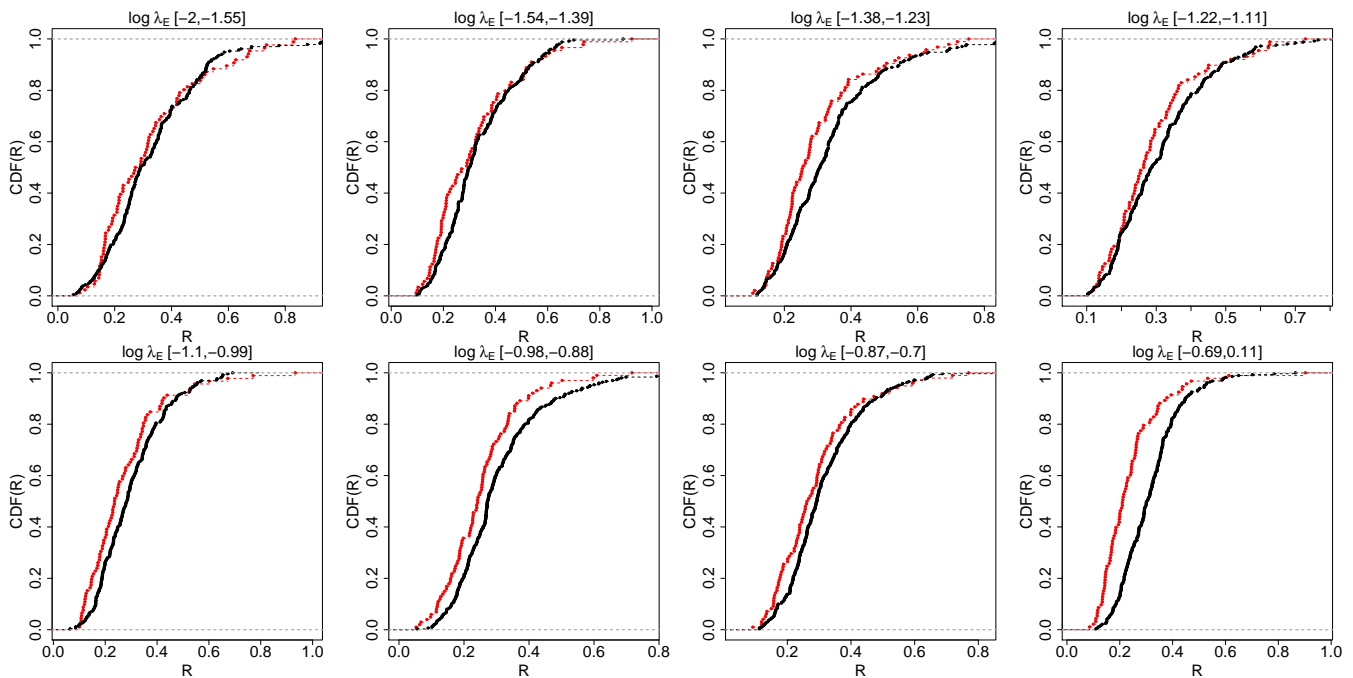}
\caption{Cumulative distributions of $R$ for the subsamples of RL quasars (\emph{red}) and RQ quasars (\emph{black}) paired with $n = 3$  for several bins of the Eddington ratio $\lambda_{\rm E}$ containing equal number of sources.}
\label{r_bin_cdf}
\end{figure*}

Figure \ref{r_bin_cdf} shows the cumulative distributions of $R$ for the samples shown in Figure \ref{best_three} divided according to $\lambda_{\rm E}$ into bins containing equal number of objects.
In order to evaluate whether there are statistical differences between the distributions of $R$ for RL and RQ samples, we make use of the two-sample Kolmogorov-Smirnov test (K-S test).
Table \ref{tbl_2} shows the $D$-values and $p$-values of each bin.
For a broad range of $\lambda_{\rm E}$ values, the values of $p$ range between 0.001 and 0.15, indicating that the difference between the distributions of $R$ are not very significant.

However, for the highest values of $\log\lambda_{\rm E} \in [-0.69:0.11]$, we obtain $p < 10^{-9}$, which indicates significant discrepancy, with the RL quasars having median covering factors lower by $\Delta R = 0.094$.

\begin{table}
\begin{center}
\caption{Tabulated results of the sample matched on z and $M_{\rm BH}$. We show the median value of $\lambda_{\rm E}$ for each of the bins, the numbers of sources $N$, the median value of $R$, the standard deviation of $R$ distribution $\sigma_{\rm R}$, and the $25^{\rm th}$ and $75^{\rm th}$ percentiles of $R$.}
 \label{tbl_3}
 \begin{tabular}{c c c c c c c} 
 \hline
 $\log\lambda_{\rm E,50}$ & class & $N$ & $R_{\rm 50}$ & $\sigma_{\rm R}$ & $R_{\rm 25}$ & $R_{\rm 75}$ \\
  \hline\hline
-1.695 & RL & 86 & 0.282  & 0.177 & 0.178 & 0.418 \\ 
-1.71  & RQ & 233 & 0.291   & 0.170 & 0.221 & 0.427 \\ 
-1.45  & RL & 89 & 0.289   & 0.164 & 0.192 & 0.391 \\ 
-1.46  & RQ & 249 & 0.301   & 0.142 & 0.234 & 0.419 \\ 
-1.31  & RL & 95 & 0.256   & 0.140 & 0.206 & 0.340 \\ 
-1.3  & RQ  & 271 & 0.307   & 0.155 & 0.226 & 0.392 \\ 
-1.17 & RL  & 88 & 0.264   & 0.132 & 0.199 & 0.343 \\ 
-1.15 & RQ  & 253 & 0.288  & 0.136 & 0.204 & 0.385 \\ 
-1.04 & RL  & 92 & 0.236   & 0.148 & 0.171 & 0.336 \\ 
-1.04 & RQ  & 261 & 0.284   & 0.127 & 0.198 & 0.371 \\ 
-0.93 & RL  & 101 & 0.242   & 0.116 & 0.178 & 0.308 \\ 
-0.93 & RQ  & 301 & 0.273   & 0.143 & 0.212 & 0.357 \\ 
-0.8 & RL   & 98 & 0.264   & 0.131 & 0.191 & 0.341 \\ 
-0.79 & RQ  & 290 & 0.293   & 0.121 & 0.234 & 0.379 \\ 
-0.54 & RL  & 93 & 0.211   & 0.122 & 0.160 & 0.268 \\ 
-0.53 & RQ  & 282 & 0.304   & 0.114 & 0.229 & 0.373 \\ 
 \hline
\end{tabular}
\end{center}
\end{table}

\begin{table}
\caption{Results of the Kolmogorov-Smirnov test applied to the distribution in $R$ of RL and paired RQ quasars binned according to the Eddington ratio $\lambda_E$.}
\label{tbl_2}
\begin{center}
 \begin{tabular}{c c c} 
 \hline
 $log \lambda_{\rm E}$ & $D$  & $p$ \\ 
 \hline\hline
$[-2.00,-1.55]$ & 0.157 & 0.096 \\
$[-1.54,-1.39]$ & 0.176 & 0.035 \\
$[-1.38,-1.23]$ & 0.193 & 0.011 \\
$[-1.22,-1.11]$ & 0.141 & 0.150 \\
$[-1.10,-0.99]$ & 0.174 & 0.033 \\
$[-0.98,-0.88]$ & 0.218 & 0.001 \\
$[-0.87,-0.70]$ & 0.148 & 0.082 \\
$[-0.69, 0.11]$ & 0.398 & 4.67x$10^{-10}$ \\
 \hline
\end{tabular}
\end{center}
\end{table}

\section{Discussion and Conclusions}
\label{sec_discussion}

While the main factor determining the radiative efficiency of accretion flows in AGN is the accretion rate normalized by the black hole mass 
(equivalently, the accretion-rate Eddington ratio $\dot M c^2/L_{\rm E}$),
the efficiency of jet production covers 3-4 orders of magnitude at any Eddington ratio \citep{2007ApJ...658..815S}.
This is particularly well documented in the case of quasars,
in which radio-loudness,
defined to be the radio-to-optical luminosity ratio \citep{1989AJ.....98.1195K},
can be monitored down to lowest values
of radio luminosity associated with jet activity.
\citep{2011ApJ...739L..29K}. 
However like in X-ray BH binaries the AGN jet activity is common only at low-rate, radiatively inefficient accretion flows.
At larger accretion rates,corresponding with $\lambda > 0.01$ at which accretion flows become radiatively efficient \citep{2014ARA&A..52..529Y},
the jet production is typically very suppressed.
This is evidenced by having only  $\sim 10\%$ of quasars associated with radio sources powered by jets,
and only about $2\%$ developing luminous, extended FRII radio structures \citep{2007ApJ...654...99W}.
In order to explain the existence of RL quasars and their rarity, 
one might consider an intermittency of jet production at high accretion rates and its low duty cycle.
Such intermittency have been proposed to be driven by stochastic switches between two accretion modes,
one driven by viscosity and another led by removing angular momentum via MHD outflows \citep{2003ApJ...593..184L,2006MNRAS.372.1366K}.
However, noting the $10^7-10^8$-year lifetimes of FRII radio sources (\citealt{2008ApJ...676..147B,2012ApJ...756..116A}, and refs. therein),
explanation for the rarity of FRII quasars in terms of the intermittent jet production scenario would require  RQ quasar phase lifetimes to be longer than the Salpeter timescale \citep{1964ApJ...140..796S}. Furthermore, spectral properties of RQ and RL quasars are strikingly similar in all bands but radio \citep{1989ApJ...347...29S,1993AJ....106..417F, 1994ApJS...95....1E,1997ApJ...475..469Z,2002ApJ...565..773T,2006AJ....131..666D,Rich06,2011ApJS..196....2S,2016ApJ...818L...1S}.
This suggests that production of powerful jets may not be related to some exceptional accretion mode,
and seems to favour powering of jets by rotating black holes \citep{1977MNRAS.179..433B}.
In such a case, the jet power scales roughly with the square of the BH spin and magnetic flux threading the BH horizon $L_{\rm j} \propto a^2\Phi_{\rm BH}$.
While value of the spin $a$ is largely determined by cosmological evolution of BHs that involves multiple accretion and merger events \citep{2013ApJ...775...94V},
the magnetic flux $\Phi_{\rm BH}$ is built up on black hole by advection of a poloidal magnetic field.
If accumulation of the magnetic flux proceeds during the quasar phase, 
one might expect stronger magnetization of the accretion flows in RL quasars than in RQ quasars.
Higher magnetization level of accretion flow should result in increased intensity of the hydromagnetic disc winds \citep{1982MNRAS.199..883B}, both within and outside the dust sublimation radius.
If we attribute the majority of AGN obscuration to dust clouds embedded in magnetized winds, the dust covering factor will be regulated by the relative power of the dusty outer wind, and the dust-free inner wind.
The presence of powerful relativistic jets in RL quasars may also affect poloidal pressure balance, pushing both the dust-free inner wind and the dusty outer wind away from the jet axis.

Until now, most studies of dust covering factor were performed  separately for RL and RQ quasars,
and focused on their dependence on luminosity, rather than on Eddington ratio (\citealt{2015ARA&A..53..365N} and refs. therein).
The only work to our knowledge which provides direct comparison of covering factors in RL and RQ quasars,
and their dependence on Eddington ratio and BH mass, is the one by \cite{2013MNRAS.430.3445M}.
No significant differences of these properties between the two quasar populations have been found.
We performed a similar study,
but in order to have a larger radio-loudness contrast between 
the RL and RQ samples,
we limited the RL quasar sample only to those associated with FRII morphology that are known to be on average much radio-louder than the entire
radio detected population of quasars \citep{2007AJ....133.1615L}. 
Furthermore, in order to avoid the potential effect of different black hole masses and source distances on the studied properties of RL versus RQ quasars,
we compared their covering factors by including only those RQ quasars,
that match the RL ones in redshift, black hole mass and Eddington ratio.

Our main results can be summarized as follows:
\begin{itemize}
\item
Median covering factors (proxied by $R$) of FR~II quasars and RQ quasars matched in $(z; M_{\rm BH}; \lambda_{\rm E})$ are comparable ($R_{\rm 50,RL} = 0.28$ and $R_{\rm 50,RQ} = 0.31$),
both having very similar, fully overlapped broad distributions, with the possible exception of the highest Eddington ratios $\log\lambda_{\rm E} > -0.7$;
\item
Dependencies of the median covering factors on the Eddington ratio, black hole mass and redshift are statistically weak.
While very similar at lowest values of $z$, $M_{\rm BH}$ and $\lambda_{\rm E}$, they diverge 
somewhat at the highest values of these parameters,
with a trend of the  median of $R_{\rm RL}$ decreasing with both $\lambda_{\rm E}$ and $M_{\rm BH}$, 
and the median of $R_{\rm RQ}$ increasing with redshift.
\end{itemize}

Very similar covering factors of dusty obscurers in FR~II and RQ quasars suggest similar
accretion conditions on parsec scales.
Furthermore, the lack of statistically significant dependence of covering factors (CF) on the Eddington ratio down to values $\lambda_{\rm E} < 0.03$ (found also by \citealt{2013MNRAS.430.3445M})
excludes the possibility that dusty obscurers could be associated with the winds powered by radiation pressure exerted on dust.
On the other hand, recent interferometric MIR observations \citep{2016A&A...591A..47L} indicate possible elongation of the dusty obscurers in the direction corresponding to the AGN polar axis (as determined by $O_{\rm III}$ line cones and/or jets), rather than in the equatorial plane.
This would indicate outflows/winds
as the loci of dust reprocessing AGN radiation into MIR.
Such outflows might be powered magnetically as suggested by \cite{1992ApJ...385..460E}, \cite{1994ApJ...434..446K} and \cite{2006ApJ...648L.101E}.
The elongated geometry of dusty obscurers is also supported by the fact that scattering cones in quasars are much narrower than predicted by torus models \citep{2016MNRAS.456.2861O}.
Significant magnetization of the accretion flows is independently supported by the fact that  contribution of magnetic fields  to vertical pressure in accretion discs can protect them against their gravitational fragmentation \citep{2007MNRAS.375.1070B,2016MNRAS.460.3488S}.
However, similar CF of dusty obscurers in RQ and RL quasars
indicate that magnetization of accretion flows in RL quasars is not stronger than in RQ quasars.
We can envisage two scenarios for reconciling this similarity with models involving the Blandford-Znajek mechanism,
according to which efficiency of jet production strongly depends on the magnetic flux.
They are:
(1) accumulation of the magnetic flux prior to the quasar phase, which may depend strongly on the environment conditions \citep{SikSta13,2013ApJ...764L..24S};
(2) magnetic flux accumulation during the quasar phase, in which case the amount of accumulated flux can on average be much larger in RL quasars if their lifetime is much longer than the lifetime of RQ quasars \citep{2015MNRAS.451.2517S}.
Obviously, noting that only few nearby AGNs were imaged by MIR interferometry,
it remains uncertain how representative are these objects for the entire AGN population.
Noting the ambiguity of models that can successfully reproduce the observed IR spectra \citep{2015ARA&A..53..365N},
better understanding of the dusty obscurer structure will need to wait for the next generation of the MIR interferometers (e.g. \citealt{2014Msngr.157....5L}).

\section*{Acknowledgements}
\addcontentsline{toc}{section}{Acknowledgements}

We thank the anonymous reviewer for helpful comments.
We acknowledge financial support by the Polish National Science Centre grants 2013/09/B/ST9/00026 and 2015/18/E/ST9/00580.



\clearpage
\newpage
\appendix
\section{Catalogue}

In Table \ref{tbl_A1} and \ref{tbl_A2}, we list the properties of our complete RL and RQ samples. The catalogues are available as supplementary material online.

\begin{table*}
\begin{center}
\caption{Description of the FR~II quasar sample catalogue.The catalogue is available as supplementary material online.}
\label{tbl_A1}

 \begin{tabular}{c c c} 
 \hline
 Column Name & Unit  & Description \\ 
 \hline\hline
SDSS\_Name &  & SDSS Name from DR7 \\
RA\_J2000 & deg &  Coordinates of quasars \\
DEC\_J2000 & deg &  Coordinates of quasars \\
z & & Redshift\\
log\_MBH & log[solar mass] & The fiducial virial BH mass \\
log\_Eddington\_ratio & & Eddington ratio based on the fiducial virial BH mass\\
log\_Lbol & log W & Bolometric luminosity\\
W3mag & mag & Magnitude of WISE W3 band\\
Flux\_W3 & $erg$ $s^{−-1}$ $cm^{−-2}$ $Hz^{−-1}$ & Flux of WISE W3 band\\
log\_v3Lv3 & log W & Luminosity  of WISE W3 band\\
LIR & log W & MIR luminosity\\
R & & Ratio of MIR luminosity to Bolometric luminosity \\
radio\_RA & deg &  R.A., geometrical centre of lobes\\
Radio\_DE & deg & Decl., geometrical centre of lobes\\
lobe\_flux & Jy & Total lobe flux \\
core\_flux & Jy &  Core flux (zero if no core is detected).\\
 \hline
\end{tabular}
\end{center}
\end{table*}

\begin{table*}
\begin{center}
\caption{Description of the RQ sample catalogue.The catalogue is available as supplementary material online.}
\label{tbl_A2}

 \begin{tabular}{c c c} 
 \hline
 Column Name & Unit  & Description \\ 
 \hline\hline
SDSS\_Name &  & SDSS Name from DR7 \\
RA\_J2000 & deg &  Coordinates of quasars \\
DEC\_J2000 & deg &  Coordinates of quasars \\
z & & Redshift\\
log\_MBH & log[solar mass] & The fiducial virial BH mass \\
log\_Eddington\_ratio & & Eddington ratio based on the fiducial virial BH mass\\
log\_Lbol & log W & Bolometric luminosity\\
W3mag & mag & Magnitude of WISE W3 band\\
Flux\_W3 & $erg$ $s^{−-1}$ $cm^{−-2}$ $Hz^{−-1}$ & Flux of WISE W3 band\\
log\_v3Lv3 & log W & Luminosity  of WISE W3 band\\
LIR & log W & MIR luminosity\\
R & & Ratio of MIR luminosity to Bolometric luminosity \\
 \hline
\end{tabular}
\end{center}
\end{table*}
\bsp	
\label{lastpage}
\end{document}